\documentclass[aps,pra,twocolumn,superscriptaddress,showpacs,showkeys,amsmath,amssymb]{revtex4-1}
\usepackage{amsmath}
\usepackage{accents}
\usepackage{bm}
\usepackage[english]{babel}
\usepackage{graphicx}
\usepackage[usenames]{color}
\usepackage{colortbl}
\usepackage{mathtools}
\usepackage{commath}
\usepackage[font=footnotesize,figurewithin=none]{caption}
\usepackage[caption=false]{subfig}
\usepackage[toc,page,title]{appendix}

\begin{document}

\title{Probing mean values and correlations of high-spin systems on a quantum computer}

\author{A. R. Kuzmak}
\email{andrijkuzmak@gmail.com}
\author{V. M. Tkachuk}
\email{voltkachuk@gmail.com}
\affiliation{Department for Theoretical Physics, Ivan Franko National University of Lviv,\\ 12 Drahomanov St., Lviv, UA-79005, Ukraine}

\date{\today}
\begin{abstract}
We consider simulation of the high spins on a quantum computer. The protocols which allow one to measure the mean value of spin and correlations between spins are proposed.
As a result, we determine the time dependence of the mean values of spin-1 in the magnetic field prepared on the ibmq-santiago quantum computer. In addition, we study the evolution of two interacting spins on the ibmq-lima quantum computer. The time-dependencies of the mean value of spin-1 and correlations between these spins are detected. Finally, we generalize these protocols for the spins of arbitrary values.
\end{abstract}

\maketitle

\section{Introduction \label{sec1}}

Richard Feynman was one of the first scientists to present the idea of computers that use quantum mechanics to provide calculations \cite{feynman1982}. He considered these computers as tools for simulating the behaviour of quantum systems. This is because the qubits, that are the heart of quantum computers, are based on the real quantum systems such as spins of atoms, superconducting qubits, photons, etc
\cite{nielsen2000,lloyd1996,loss1998,kane1998}.
Depending on the dimensionality of the Hilbert space of a certain quantum system, the appropriate number of qubits is used
to simulate its evolution \cite{OCGQC,GAQCLB,QCAG,QGDM,GQCQ}. In the case of many-body systems, quantum computers in comparison with the classical ones provide exponential supremacy in time.
Using the Google quantum computer such a supremacy was achieved \cite{arute2019}. The authors spent about 200 seconds, instead of 10 000 years, to prepare and measure the 53-qubit quantum state. However,
the scientists from IBM argue that this problem can be solved on classical computers in 2.5 days with far greater fidelity \cite{pednault2019}. The supremacy of quantum computers is achieved due to the
nature of qubits which can be in superposition of all possible states. For example, the Hilbert space of the 53 spin-1/2 system is defined by $2^{53}$ basis states. To simulate the quantum evolution of such a system
the classical computers use $2^{53}$ bits. The quantum computers solve this problem using only 53 qubits. Moreover, each of these qubits is a two-level quantum system
which naturally allows one to simulate spin-1/2.

There are many algorithms that allow one to study the properties of different systems simulated on quantum computers. In papers \cite{wang2018,mooney2019} the
protocol that allows one to measure the negativity between pairs of qubits on a quantum computer was proposed. This protocol was applied to graph states prepared on IBM's quantum computers.
Another method allows us to determine the geometry measure of entanglement by the mean value of spin \cite{Frydryszak2017} in the case of pure \cite{kuzmak2020} and mixed \cite{kuzmak2021} quantum states prepared on the quantum computers. This method was used for measurement of the value of entanglement of the Schr\"odinger cat state \cite{kuzmak20202} and graph states of spin system generated by the Ising
interaction \cite{gnatenko2021,gnatenko20212} prepared on the IBM's quantum computers. One of the most important problems which can be solved by quantum computers is the determination of the energy levels of quantum systems.
One of the first algorithms for the determination of the energy levels on quantum computers is based on the quantum phase estimation \cite{abrams1997,abrams1999,kitaev1997}.
In paper \cite{Cervera2018}, the author developed the protocol that diagonalizes the Hamiltonian and gives all eigenstates of the model.
Using this protocol, an evolution of the one-dimensional transverse Ising spin model was simulated on a quantum computer.
Recently, it was proposed a method for determining energy levels of quantum systems on a quantum computer \cite{gnatenko2022,gnatenko20222}.
This method is based on the study of the mean value of a physical quantity when its operator anticommutes with the Hamiltonian of a system.
It was also proposed protocols that allow one to determine other properties of quantum systems. Using the Suzuki-Trotter decomposition \cite{trotter1959,suzuki2005}, the correlations between
spins $1/2$ described by the Heisenberg model were studied on quantum simulators in papers \cite{tacchino2019,chiesa2019}.
The geometry of the states prepared on the quantum computers was studied in paper \cite{kuzmak20212}. The spin-1 tunneling was studied both on a digital simulator \cite{santini2011,chiesa2015}
and a quantum computer \cite{gnatenko20223}. In paper \cite{gnatenko20223} splitting energy levels as a result of tunneling were additionally considered.

All the above-mentioned methods are mainly related to the study of quantum systems consisting of spin-1/2 particles. However, there are many quantum systems such as
nuclei of different atoms or molecular compounds that consist of high spins. It is important to find the methods for simulation the behaviour of such systems on a quantum computer.
We propose methods that allow one to study the evolution of high-spin systems performed on quantum computers.

In this paper, we consider the method to simulate the high spins using the qubits of quantum computers. We propose the protocols to measure the mean values of spin-1 and correlations between two spins (Sec.~\ref{spin1}).
In Sec.~\ref{qcsantiago}, we implement these protocols on the IBM quantum computers to study the evolution of spin-1 in the magnetic field (Subsec.~\ref{spinmagf})
and evolution of two interacting spins described by the Ising model (Subsec.~\ref{Isingmodel}). We generalize these protocols to a spin of an arbitrary value (Sec.~\ref{spins}). Conclusions are presented in Sec.~\ref{conc}.

\section{Simulating a spin-1 on a quantum computer \label{spin1}}

The operators of spin-$s$ ${\bf S}$ and its projection on the $z$-axis $S^z$ satisfies the following equations
\begin{eqnarray}
&&{\bf S}^2\vert s, m\rangle = s(s+1)\vert s, m\rangle,\label{condition1}\\
&&S^z\vert s, m\rangle=m\vert s, m\rangle,
\label{condition2}
\end{eqnarray}
where $\vert s, m\rangle$ is the eigenstates of $S^z$ operator which correspond to the eigenvalues $-s\leq m\leq s$. We put $\hbar=1$, which means that the energy is measured in the frequency units.
Since the quantum computer consists of two-level quantum systems (qubits), to simulate the high spin we use a set of spins $1/2$. To simulate the spin of value $1$ it is sufficient to take two spins $1/2$.
Then the spin-1 operator can be expressed as the sum of two spin-1/2 operators
\begin{eqnarray}
{\bf S}=\frac{1}{2}\left({\boldsymbol \sigma}_1+{\boldsymbol \sigma}_2\right),
\label{spinoperator1}
\end{eqnarray}
where we introduce the following notations ${\boldsymbol \sigma_1}={\boldsymbol \sigma}\otimes I$, ${\boldsymbol \sigma_2}=I\otimes{\boldsymbol \sigma}$ which mean that the Pauli operator ${\boldsymbol \sigma}$
acts on the $1$st, $2$nd spin-$1/2$ while the identity operator $I$ acts on the $2$nd, $1$st spin-$1/2$, respectively. Here
${\boldsymbol \sigma}=\sigma^x{\bf i}+\sigma^y{\bf j}+\sigma^x{\bf k}$ is the Pauli operator presented in the Cartesian coordinate system.
It is easy to see that the components of this operator satisfy the Lie algebra $\left[S^{\alpha},S^{\beta}\right]=i\epsilon_{\alpha\beta\gamma}S^{\gamma}$,
where $\epsilon_{\alpha\beta\gamma}$ is the Levi-Civita symbol. Substituting operator (\ref{spinoperator1}) into equations (\ref{condition1}), (\ref{condition2})
we obtain
\begin{eqnarray}
&&\frac{1}{4}\left(6I+2\left(\sigma^x_1\sigma^x_2+\sigma^y_1\sigma^y_2+\sigma^z_1\sigma^z_2\right)\right)\vert 1, m\rangle = 2\vert 1, m\rangle,\label{condition1spin1}\\
&&\frac{1}{2}\left(\sigma_1^z+\sigma_2^z\right)\vert 1, m\rangle=m\vert 1, m\rangle.
\label{condition2spin1}
\end{eqnarray}
Solving these equations, we obtain the states that allow one to determine the Hilbert space of spin-1 \cite{Sinha2003}
\begin{eqnarray}
&&\vert 1, 1\rangle= \vert 00\rangle,\label{szeigenstatessspin10}\\
&&\vert 1, 0\rangle= \frac{1}{\sqrt{2}}\left(\vert 01\rangle+\vert 10\rangle\right),\label{szeigenstatessspin101}\\
&&\vert 1, -1\rangle= \vert 11\rangle,
\label{szeigenstatesspin1}
\end{eqnarray}
where $\vert 0\rangle$, $\vert 1\rangle$ are the eigenstates of $\sigma^z$ operator which correspond to +1, -1 eigenvalues, respectively.  An arbitrary state of spin-$1$ can be expressed as the decomposition of these
eigenstates
\begin{eqnarray}
\vert\psi_1\rangle=C_1\vert 1, 1\rangle+C_0\vert 1, 0\rangle+C_{-1}\vert 1, -1\rangle,
\label{arbstatespin1}
\end{eqnarray}
where $C_m$ are the complex parameters which satisfy the normalization condition $\sum_{m=-1}^1\vert C_m\vert^2=1$. Note that the operator (\ref{spinoperator1}) does not extend state (\ref{arbstatespin1})
beyond the Hilbert space spanned by basis vectors (\ref{szeigenstatessspin10})-(\ref{szeigenstatesspin1}).

Let us summarize all the above. Components of the operator (\ref{spinoperator1})  provide the transformations of the states on the subspace spanned by states (\ref{szeigenstatessspin10}), (\ref{szeigenstatessspin101}), and (\ref{szeigenstatesspin1})
in the same way as components of the single spin-1 operator act on its Hilbert space. Another important point is that the operator (\ref{spinoperator1}) with these states satisfies conditions (\ref{condition1}), and (\ref{condition2})
for $s = 1$. The components of the operator (\ref{spinoperator1}) also satisfy the Lie algebra for the spin-1 operator. The components of this operator do not extend states beyond the Hilbert space spanned by states
(\ref{szeigenstatessspin10}), (\ref{szeigenstatessspin101}), and (\ref{szeigenstatesspin1}). This means that during evolution the state does not extend beyond this Hilbert space. These conditions allow us
to represent spin-1 by two spin-1/2. This idea is not new. It is described in paper \cite{Sinha2003}. Based on this idea we propose a simulation of spin-1 on a quantum computer.
In addition, in some way, we propose a simulation of an arbitrary high-spin by a certain number of spins 1/2.

The quantum computer performs measurements on a basis $\vert 00\rangle$,
$\vert 01\rangle$, $\vert 10\rangle$,  $\vert 11\rangle$, and as a result gives us the probabilities $P_{00}$, $P_{01}$, $P_{10}$, $P_{11}$. These values are related with squares of modules of amplitudes which are included by state (\ref{arbstatespin1}) in the following way
\begin{eqnarray}
&&\vert C_1\vert^2=P_{00},\nonumber\\
&&\vert C_0\vert^2=P_{01}+P_{10},\nonumber\\
&&\vert C_{-1}\vert^2=P_{11}.
\label{relationsampl}
\end{eqnarray}

\begin{figure}[!!h]
\includegraphics[scale=0.6, angle=0.0, clip]{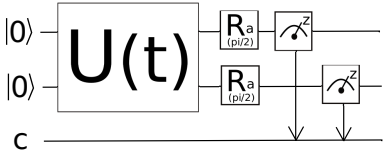}
\caption{Quantum circuit for measuring the mean value of
$S^{\alpha}$ component of spin-1 operator (\ref{meansspin1}) in the state generated
by the unitary operator $U(t)$. Depending on the spin component the $R_a$ gate provides the rotation of the spin-1 state around the $a$-axis by angle $\pi/2$.}
\label{protocolspin1}
\end{figure}

We propose to study the evolution of spin-1 described by a some Hamiltonian $H$ on a quantum computer. For this purpose we consider the evolution of mean value of spin
\begin{eqnarray}
\langle\psi(t)\vert{\bf S}\vert\psi(t)\rangle=\langle\psi_I\vert e^{iHt}{\bf S}e^{-iHt}\vert\psi_I\rangle,
\label{meanvaluespin1}
\end{eqnarray}
where $\vert\psi(t)\rangle=e^{-iHt}\vert\psi_I\rangle$ is the state achieved during the evolution having started from the initial state $\vert\psi_I\rangle$. The unitary operator $U(t)=e^{-iHt}$ is implemented
by applying a sequence of gates of a quantum computer. In papers \cite{kuzmak2020,kuzmak20202,gnatenko2021,kuzmak2021}, the protocols for measuring the mean value of spin-$1/2$ on a quantum computer are considered.
We propose a protocol that allows us to measure the evolution of the mean value of spin-1 on a quantum computer. Using the fact that in general quantum computers provide measurements of each spin on the $z$-direction we represent the $S^x$ and $S^y$ components
of spin-1 operator in the form
\begin{eqnarray}
S^x=e^{-i\frac{\pi}{2}S^y}S^ze^{i\frac{\pi}{2}S^y},\quad S^y=e^{i\frac{\pi}{2}S^x}S^ze^{-i\frac{\pi}{2}S^x}.
\label{representationofspinoperator}
\end{eqnarray}
Then, using the fact that $S^z$ operator can be decomposed by eigenstates (\ref{szeigenstatessspin10})-(\ref{szeigenstatesspin1}) as follows
\begin{eqnarray}
S_z=1\vert 1, 1\rangle\langle 1, 1\vert + 0\vert 1, 0\rangle\langle 1, 0\vert-1\vert 1, -1\rangle\langle 1, -1\vert,
\label{zcompofspin1}
\end{eqnarray}
the means take the form
\begin{eqnarray}
&&\langle\psi(t)\vert S^x\vert\psi(t)\rangle=\langle\psi(t)\vert e^{-i\frac{\pi}{2}S^y}S^ze^{i\frac{\pi}{2}S^y}\vert\psi(t)\rangle\nonumber\\
&&=\vert\langle\tilde{\psi}^y(t)\vert 1,1\rangle\vert^2-\vert\langle\tilde{\psi}^y(t)\vert 1,-1\rangle\vert^2,\nonumber\\
&&\langle\psi(t)\vert S^y\vert\psi(t)\rangle=\langle\psi(t)\vert e^{i\frac{\pi}{2}S^x}S^ze^{-i\frac{\pi}{2}S^x}\vert\psi(t)\rangle\nonumber\\
&&=\vert\langle\tilde{\psi}^x(t)\vert 1,1\rangle\vert^2-\vert\langle\tilde{\psi}^x(t)\vert 1,-1\rangle\vert^2,\nonumber\\
&&\langle\psi(t)\vert S^z\vert\psi(t)\rangle=\vert\langle\psi(t)\vert 1,1\rangle\vert^2-\vert\langle\psi(t)\vert 1,-1\rangle\vert^2,
\label{meansspin1}
\end{eqnarray}
where $\vert\tilde{\psi}^x(t)\rangle=e^{-i\frac{\pi}{2}S^x}\vert\psi(t)\rangle$, $\vert\tilde{\psi}^y(t)\rangle=e^{i\frac{\pi}{2}S^y}\vert\psi(t)\rangle$. The mean values contain the compositions of probabilities
that define the projections of spin-1 on the $\vert 1,1\rangle$, $\vert 1,-1\rangle$ states. As we can see, before measuring the mean values of $S^x$ and $S^y$ operators, the spin should be rotated around the $y$- and
$x$-axes by the angles $\pi/2$, respectively. The protocol that allows one to determine the means of spin-1 operator on a quantum computer is presented in Fig.~\ref{protocolspin1}.

In the system which consists of a set of interacting spin-1, in addition to measuring the time-dependence of the mean value of spin, we can study the evolution of correlation functions between certain spins.
Using definitions (\ref{representationofspinoperator}), (\ref{zcompofspin1}) the correlation function between two spins can be expressed as follows
\begin{eqnarray}
&&\langle\psi(t)\vert S^{\alpha}_iS^{\beta}_j\vert\psi(t)\rangle=\langle\tilde{\psi}(t)\vert S^z_iS^z_j \vert\tilde{\psi}(t)\rangle\nonumber\\
&&=\vert\langle\tilde{\psi}(t)\vert\vert 1,1\rangle_i\vert 1,1\rangle_j\vert^2-\vert\langle\tilde{\psi}(t)\vert\vert 1,1\rangle_i\vert 1,-1\rangle_j\vert^2\nonumber\\
&&-\vert\langle\tilde{\psi}(t)\vert\vert 1,-1\rangle_i\vert 1,1\rangle_j\vert^2+\vert\langle\tilde{\psi}(t)\vert\vert 1,-1\rangle_i\vert 1,-1\rangle_j\vert^2,
\label{correlatfunc}
\end{eqnarray}
\begin{figure}[!!h]
\includegraphics[scale=0.5, angle=0.0, clip]{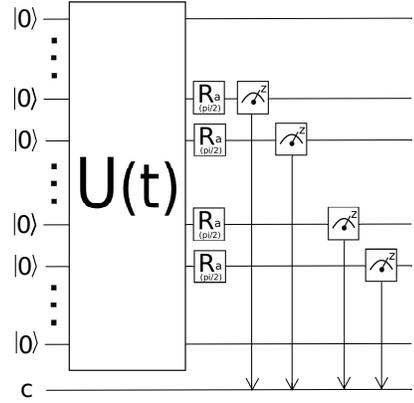}
\caption{Quantum circuit for measuring the correlation between pair of spin-1
(\ref{correlatfunc}) in the state generated by the unitary operator $U(t)$. Depending on the spin component the $R_a$ gate provides the rotation of the spin-1 state around the $a$-axis by angle $\pi/2$.}
\label{protocolspin2}
\end{figure}
where $\vert\tilde{\psi}(t)\rangle$ is the wave function of the system after corresponding rotations defined by equation (\ref{representationofspinoperator}). To determine the correlation function between two spins-1
The probabilities of their projections on four basis states should be measured. The protocol that allows one to determine the correlation on a quantum computer is presented in Fig.~\ref{protocolspin2}.
Let us apply these protocols to determine the evolution of spin-1 systems prepared on the quantum computers.

\section{Implementation of the evolution of spin-1 systems on the ibm quantum computers \label{qcsantiago}}

\begin{figure}[!!h]
\subfloat[]{\label{ibmq-s}}\includegraphics[scale=0.4, angle=0.0, clip]{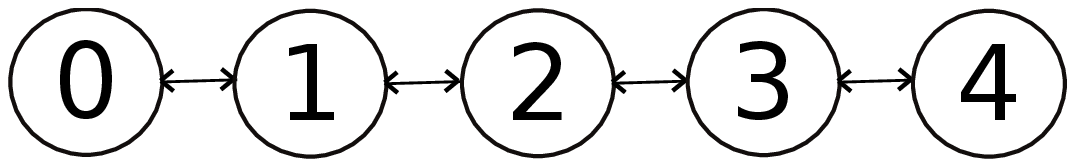}\qquad
\subfloat[]{\label{ibmq-l}}\includegraphics[scale=0.3, angle=0.0, clip]{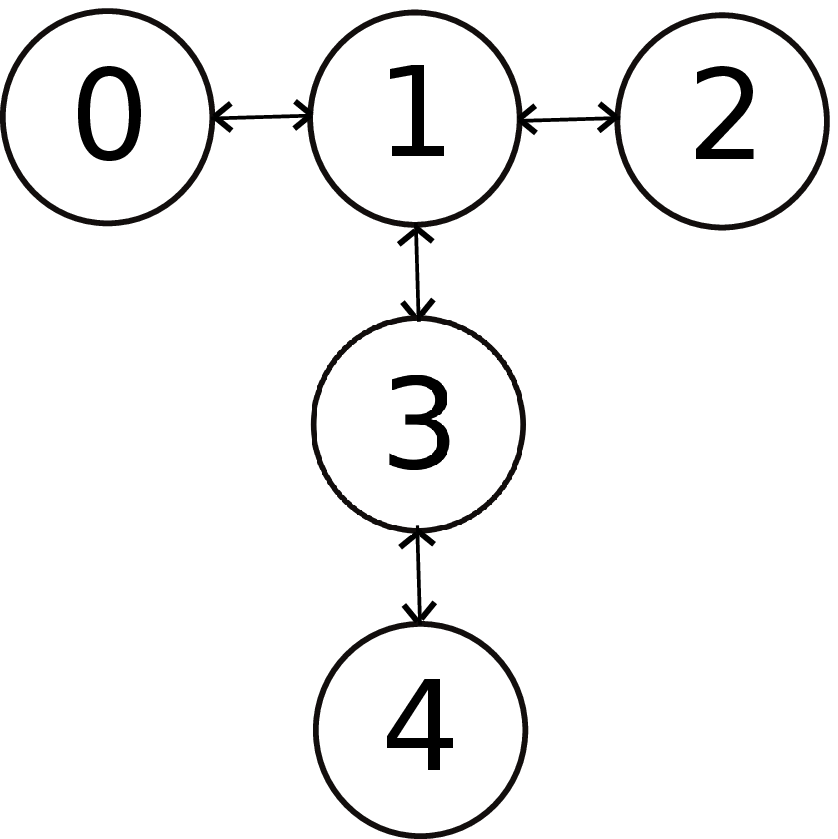}
\caption{Scheme of connection of qubits in the ibmq-santiago (a) and ibmq-lima (b) quantum devices. The bidirectional arrow defines the application of a controlled-NOT gate to a certain pair of qubits.
	Each of the pair can be controlled by this gate in a way that each of the qubits can be both controlled and target.}
\label{ibmq_santiago}
\end{figure}

We test our protocol on two quantum computers called ibmq-santiago and ibmq-lima. These devices have five qubits quantum processors each of which can be used freely through the IBM Q Experience cloud service \cite{IBMQExp}. These devices
consist of five superconducting qubits and have the structures presented in Fig.~\ref{ibmq_santiago}. An arbitrary quantum circuit can be performed by these computers using the following quantum gates:
controlled-NOT gate, the identity gate, $\sigma^x$ Pauli operator, $\sqrt{\sigma^x}$ gate and the $R_z(\phi)$ gate which corresponds to rotating the qubit around the $z$-axis by the angle $\phi$ \cite{OpenQasm}.

In the following subsection, on the ibmq-santiago quantum computer, we study the evolution of the mean value of spin-1. In the last subsection, on the ibmq-lima quantum computer, we consider the evolution of two spins
described by the Ising model.

\subsection{Spin-1 in the magnetic field \label{spinmagf}}

The Hamiltonian of spin-$1$ in the magnetic field has the form
\begin{eqnarray}
H=\omega {\bf S}\cdot {\bf n},
\label{hamspin1}
\end{eqnarray}
where ${\bf S}$ is defined by expression (\ref{spinoperator1}), $\omega$ defines the value of interaction between the spin and magnetic field, and ${\bf n}=\left(n_x,n_y,n_z\right)$
is the unit vector that determines the direction of the magnetic field. The operator of evolution with the Hamiltonian takes the form
\begin{eqnarray}
&&U=\left(\cos\left(\frac{\omega t}{2}\right)I-i\sin\left(\frac{\omega t}{2}\right) {\boldsymbol \sigma}\cdot {\bf n}\right)\nonumber\\
&&\otimes\left(\cos\left(\frac{\omega t}{2}\right)I-i\sin\left(\frac{\omega t}{2}\right) {\boldsymbol \sigma}\cdot {\bf n}\right).
\label{evolutionoperatorspin1}
\end{eqnarray}
In the case of the perpendicular orientation of the magnetic field with respect to the state, evolution happens
with the maximal speed \cite{boscain2006,frydryszak2008,boozer2012,Tkachuk2011,brachass}. Let us consider the evolution of spin-1 in the perpendicular orientation of the magnetic field
having started from the states projected on the $z$-axis. On IBM's quantum computers, this evolution can be implemented by applying to each qubit the $U(\theta,\phi,\lambda)$,
where $\theta$, $\phi$ and $\lambda$ are real parameters that can take the values from range $[0,2\pi]$. This gate is represented by the following gates of the quantum computer
\begin{figure}[!!h]
\subfloat[]{\label{mean_amplitudes_spin1_mf}}\includegraphics[scale=0.58, angle=0.0, clip]{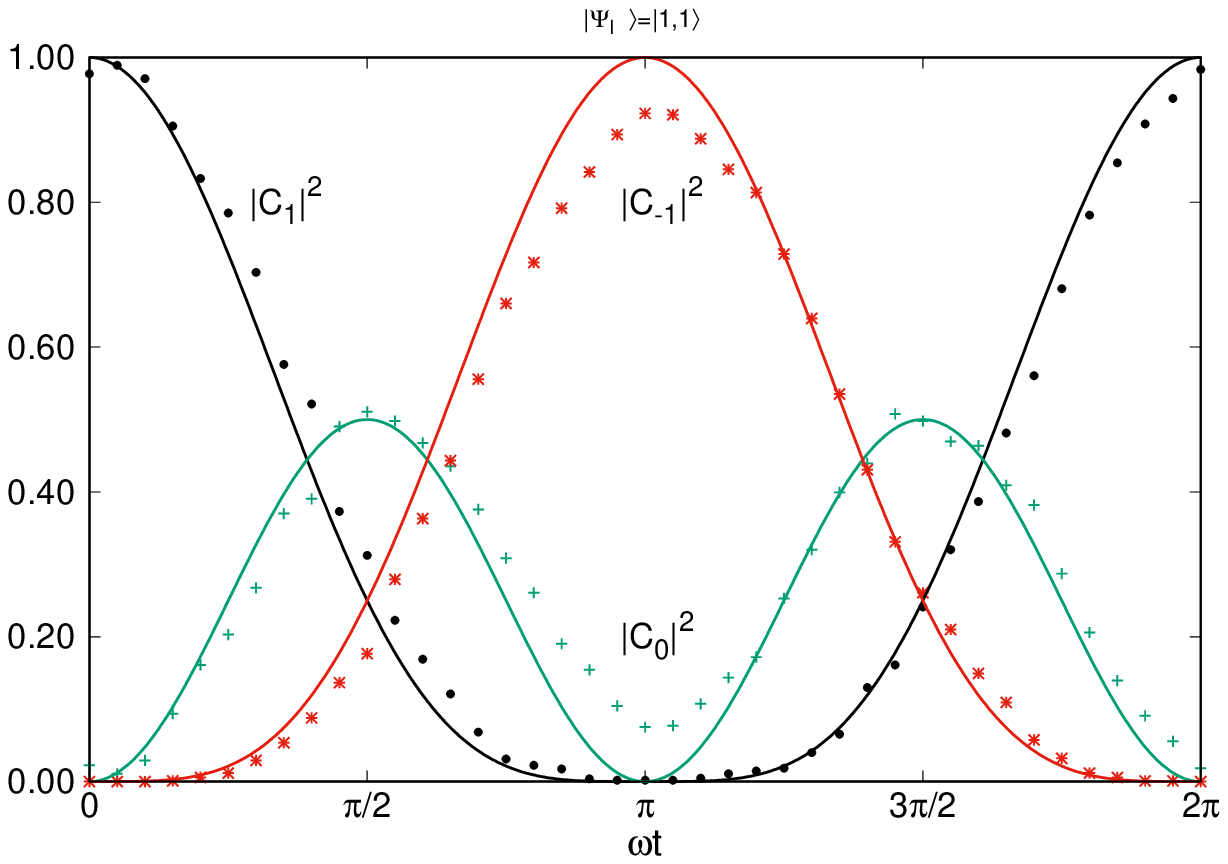}
\subfloat[]{\label{mean_amplitudes_spin1_mf2}}\includegraphics[scale=0.58, angle=0.0, clip]{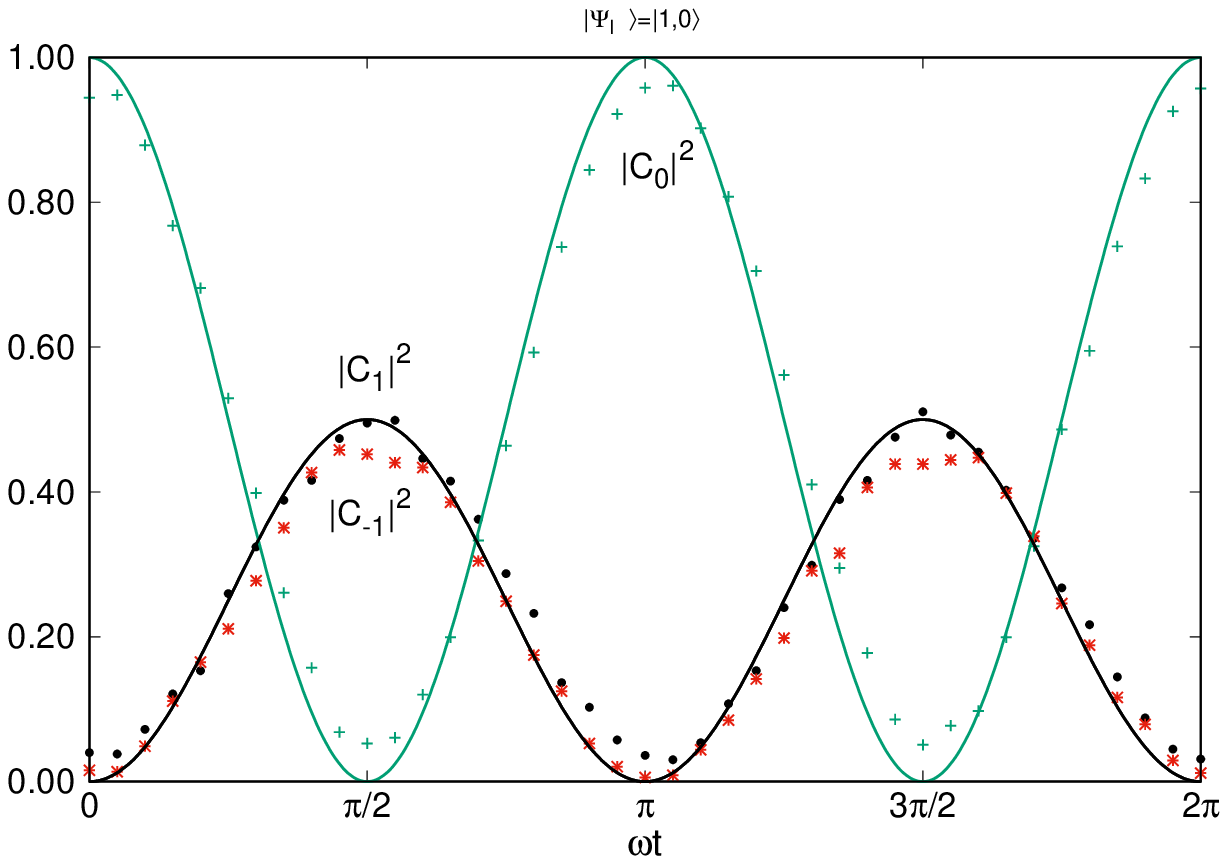}
\caption{The evolution of square of module amplitudes of spin-1 state in the magnetic field directed along the $x$-axis when the initial states are defined by expressions (\ref{szeigenstatessspin10}) and (\ref{szeigenstatessspin101}), respectively.
The solid lines show theoretical prediction and the dots present results obtained on the ibmq-santiago quantum computer.}
\label{amplitudes}
\end{figure}
\begin{eqnarray}
&&U(\theta,\phi,\lambda)\nonumber\\
&&=R_z(\phi+\pi)\sqrt{\sigma^x}R_z(\theta-\pi)\sqrt{\sigma^x}R_z(\lambda).
\label{U3gatebasis}
\end{eqnarray}
In the basis $\vert 0\rangle$, $\vert 1\rangle$ this gate reads
\begin{eqnarray}
U(\theta,\phi,\lambda)=\left( \begin{array}{ccccc}
\cos\frac{\theta}{2} & -e^{i\lambda}\sin{\frac{\theta} {2}} \\
e^{i\phi}\sin{\frac{\theta} {2}} & e^{i\left(\lambda+\phi\right)}\cos\frac{\theta}{2}
\end{array}\right).
\label{U3gate}
\end{eqnarray}
Thus, in the case of the perpendicular orientation of the magnetic field, the gate parameters are compared with the evolution parameters as follows: $\theta=\omega t$, $\phi=\arctan\left(\frac{n_y}{n_x}\right)-\frac{\pi}{2}$
and $\lambda=-\arctan\left(\frac{n_y}{n_x}\right)+\frac{\pi}{2}$.

\begin{figure}[!!h]
\subfloat[]{\label{mean_val_spin1_mf}}\includegraphics[scale=0.58, angle=0.0, clip]{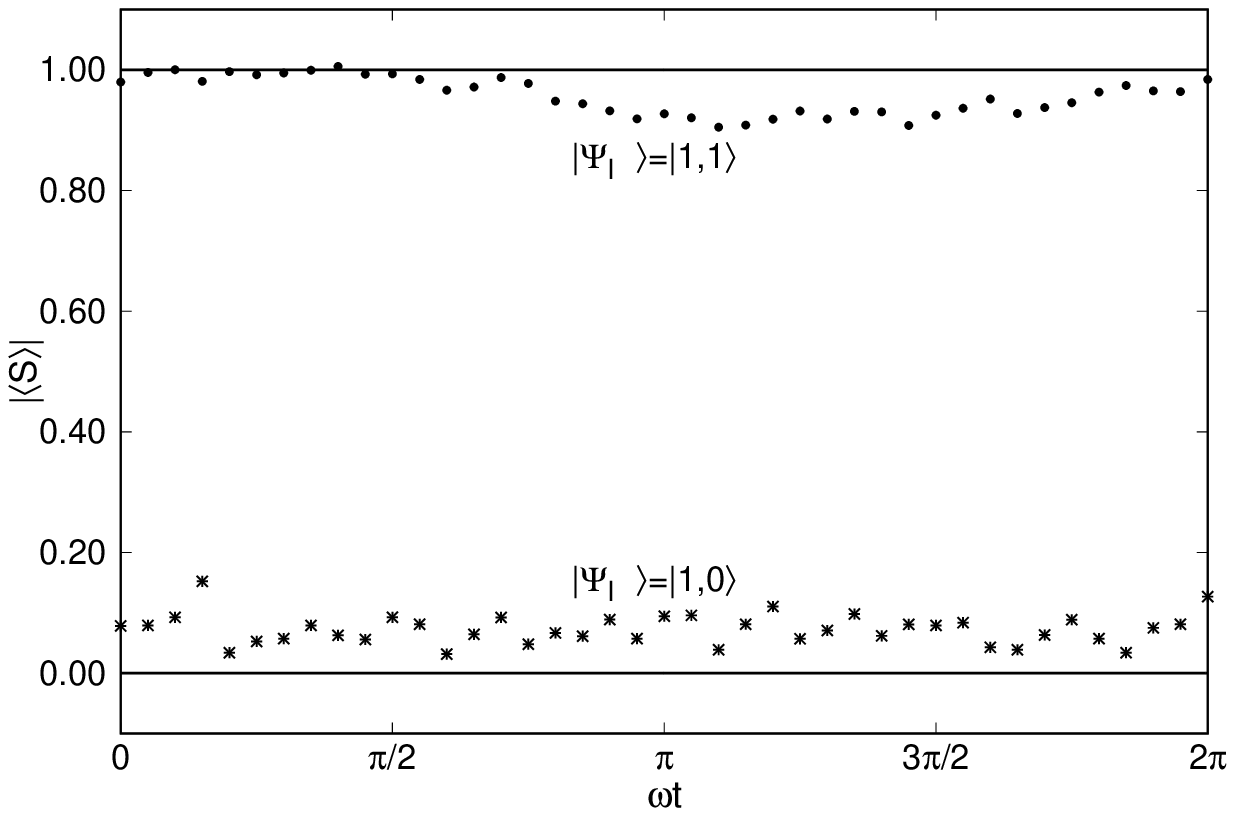}
\subfloat[]{\label{mean_sz_spin1_mf2}}\includegraphics[scale=0.58, angle=0.0, clip]{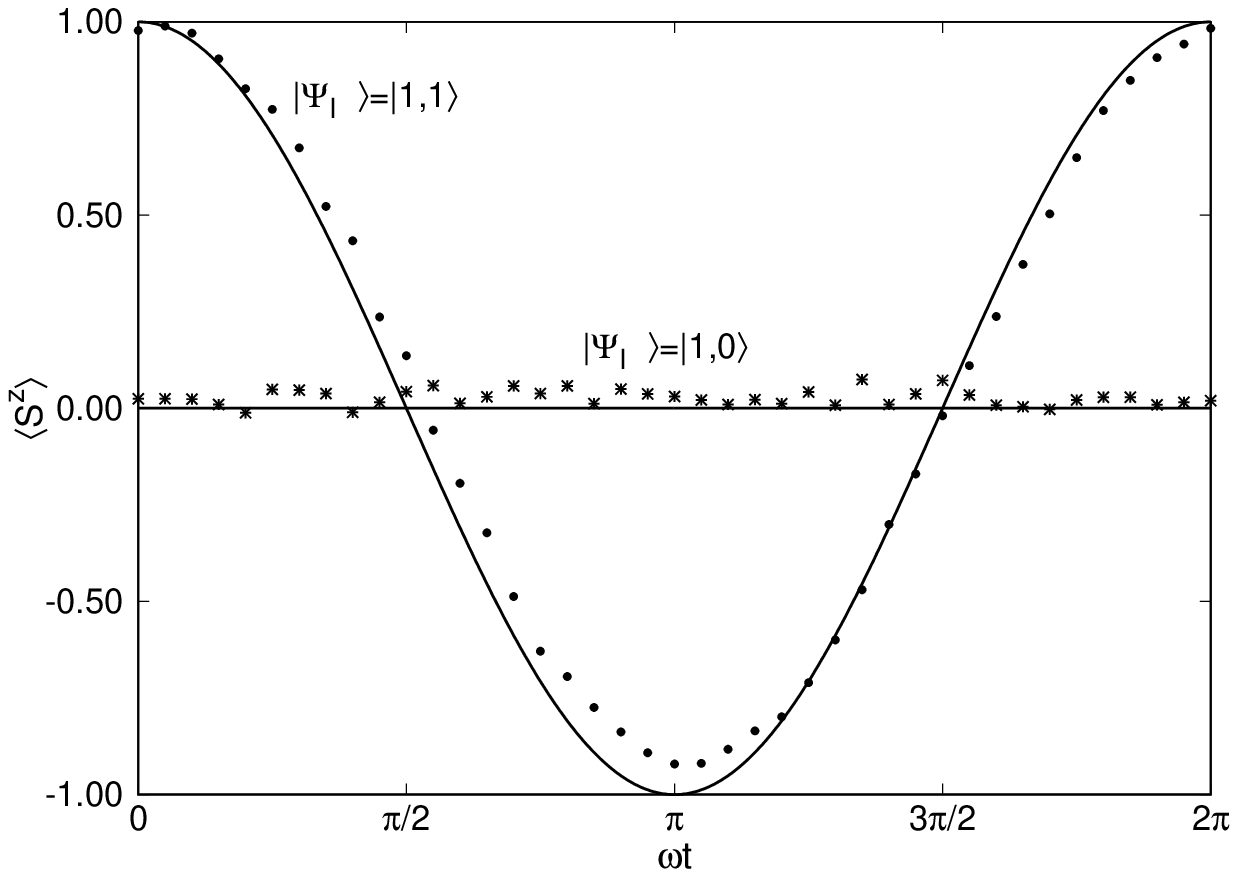}
\caption{The evolution of the mean value of spin-1 operator (a) and mean value of $S^z$ operator (b) in the magnetic field directed along the $x$-axis when the initial states are defined by equations (\ref{szeigenstatessspin10}) and (\ref{szeigenstatessspin101}), respectively.
The solid lines show theoretical prediction and the dots present results obtained on the ibmq-santiago quantum computer.}
\label{meanvalspin1}
\end{figure}

We consider the case when the magnetic field is directed along the $x$-axis. The operator that defines such an evolution has the form (\ref{evolutionoperatorspin1}) with $n=(1,0,0)$.
On the quantum computer, we change $\omega t$ in the range from 0 to $2\pi$ with step $\pi/20$. Thus for each value of $\omega t$ the means of spin components (\ref{meansspin1}) are calculated.
For each of these components, the quantum computer measures 1024 times. In Fig.~\ref{amplitudes}
we present the evolution of the square of module amplitudes in the case of two initial states (\ref{szeigenstatessspin10}) and (\ref{szeigenstatessspin101}), respectively. These values are the probabilities of the projection of the spin-1 state on the basis states
(\ref{szeigenstatessspin10}), (\ref{szeigenstatessspin101}) and (\ref{szeigenstatesspin1}), respectively. Thus, in the case when evolution begins from the state $\vert 1,1\rangle$ these probabilities evolve as follows
$\vert C_1\vert^2=\cos^4\frac{\omega t}{2}$, $\vert C_0\vert^2=\frac{1}{2}\sin^2\omega t$, $\vert C_{-1}\vert^2=\sin^4\frac{\omega t}{2}$ (Fig.~\ref{mean_amplitudes_spin1_mf}).
In the case when evolution begins from the state $\vert 1,0\rangle$ probabilities are determined by expressions $\vert C_1\vert^2=\vert C_{-1}\vert^2=\frac{1}{2}\sin^2\omega t$, $\vert C_0\vert^2=\cos^2\omega t$
(Fig.~\ref{mean_amplitudes_spin1_mf2}). Also, we measure the mean value of spin-1 in the magnetic field. In the case of $\vert\psi_I\rangle=\vert 1,1\rangle$ the mean values of spin components
have the following form $\langle S^x\rangle=0$, $\langle S^y\rangle=-\sin\omega t$, $\langle S^z\rangle=\cos\omega t$. In the case of $\vert\psi_I\rangle=\vert 1,0\rangle$ all mean values of spin components
are equal zero $\langle S^x\rangle=\langle S^y\rangle=\langle S^z\rangle=0$. In Fig.~\ref{meanvalspin1} we show time evolution in the magnetic field of the mean value of spin-1 and the mean value of the $z$-component of spin-1.

Due to the gate and readout errors that appear on the quantum computers the results obtained on the quantum computers do not exactly coincide with theoretical predictions. In addition, these errors
change the quantum state in the basis (\ref{szeigenstatesspin1}) they also take this state beyond the basis. Thus, the term with the singlet state $\frac{1}{\sqrt{2}}\left(\vert 01\rangle-\vert 10\rangle\right)$
appears in the measurements as an error. It is possible to improve the results if we provide the measurements on the basis spanned by (\ref{szeigenstatesspin1}) and singlet states. Then we can take the results of measurements
that coincide with (\ref{szeigenstatesspin1}) states and discard those that coincide with the singlet state. However, such measurements require additional applications of the corresponding unitary operator,
which adds new gate errors.

\subsection{Ising model \label{Isingmodel}}

To show that the method works, we consider the simplest circuit that can be implemented with the presence of spin-1. For this purpose, we study the evolution
of spin-1 which interacts with spin-1/2 by the Ising model. The Hamiltonian of the model has the form
\begin{eqnarray}
H=JS_1^z\sigma_2^z,
\label{HamIM}
\end{eqnarray}
where $J$ is the coupling constant between spins. In this case, three qubits should be used to simulate the system on a quantum computer. Thus, the operators of spin-$1$ and spin-$1/2$ in the Hamiltonian have the following form
$S_1^z=\frac{1}{2}\left(\sigma_z\otimes I\otimes I+I\otimes\sigma_z\otimes I\right)$ and $\sigma_2^z=I\otimes I\otimes\sigma^z$, respectively. The evolution of this system can be expressed as follows
\begin{eqnarray}
\vert\psi(t)\rangle=e^{-iJt S_1^z\sigma_2^z}\vert\psi_I\rangle.
\label{evolutIM}
\end{eqnarray}
\begin{figure}[!!h]
\includegraphics[scale=1.30, angle=0.0, clip]{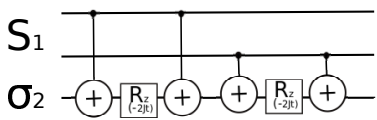}
\caption{Quantum circuit which allows one to realize evolution between two interacting spins on a quantum computer (\ref{evolutIM}). First and second pairs of qubits define $i$th and $j$th spins, respectively.}
\label{schemeIM}
\end{figure}
\begin{figure}[!!h]
\subfloat[]{\label{mean_val_spin1_IM}}\includegraphics[scale=0.58, angle=0.0, clip]{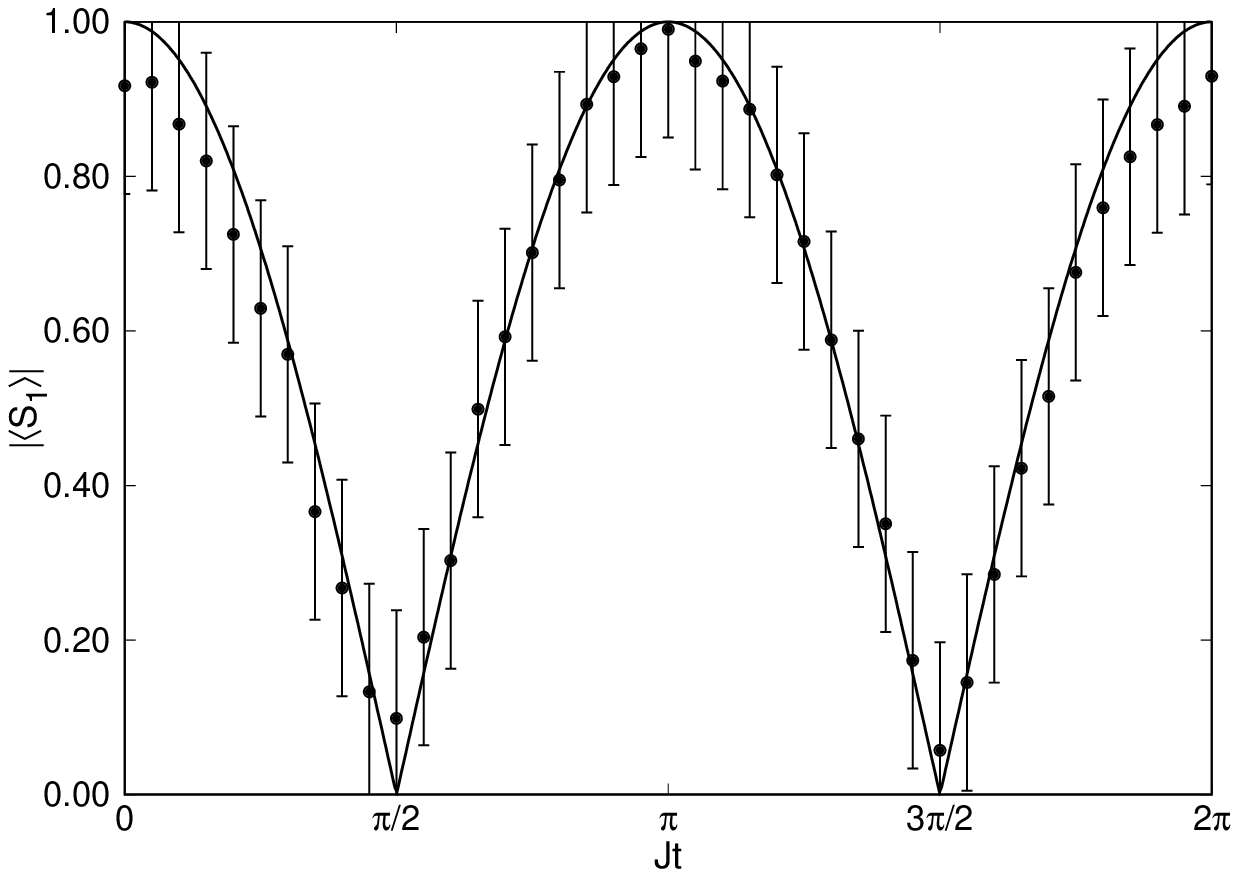}
\subfloat[]{\label{corrIM}}\includegraphics[scale=0.58, angle=0.0, clip]{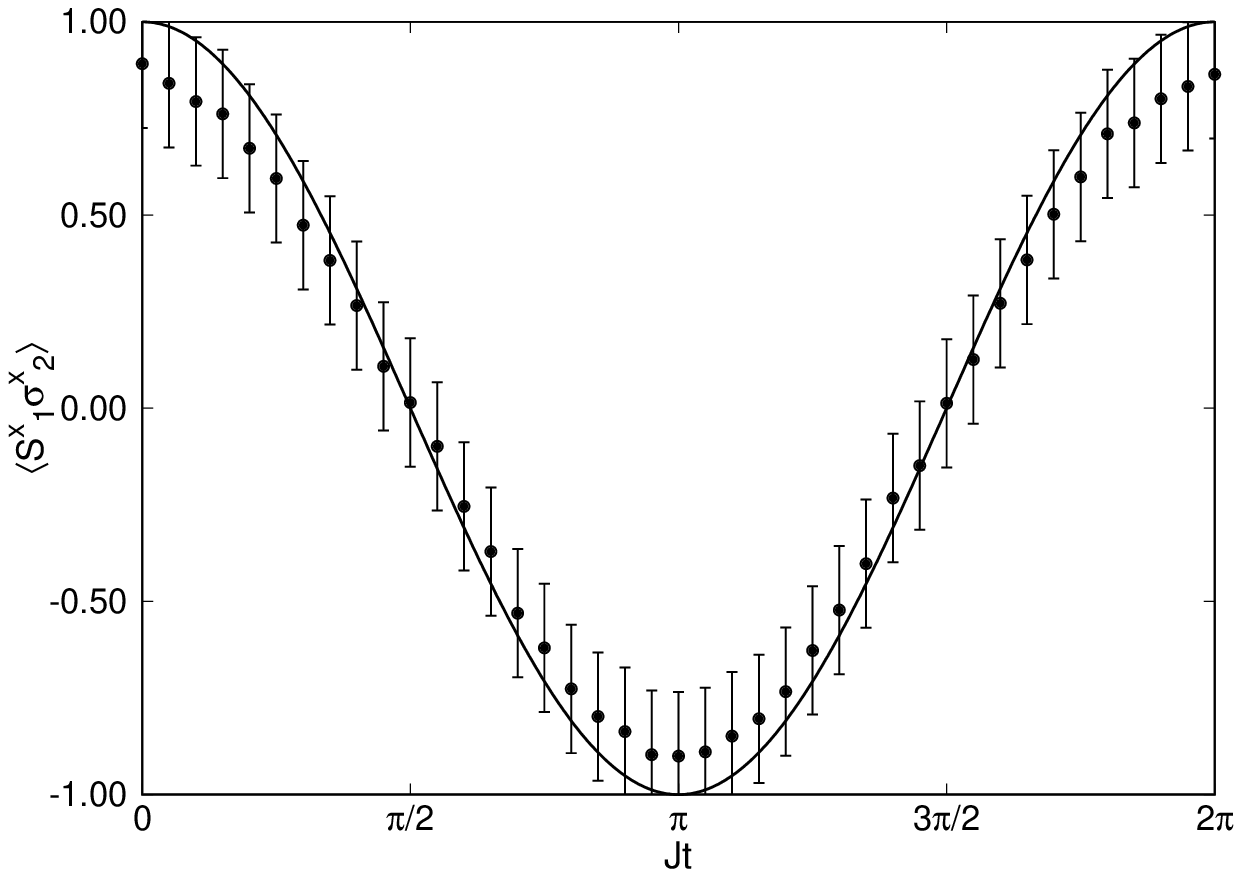}
\caption{The evolution of the mean value of spin-1 (a) and the correlation between spin-1 and spin-1/2 are described by the Ising model. Solid lines show theoretical prediction and dots present results obtained on ibmq-lima quantum computer.}
\label{meanvalspin1corrIM}
\end{figure}
The circuit which describes this evolution is shown
in Fig.~\ref{schemeIM}. It consists of two segments that define the Ising interaction between first, and third qubits and second, and third qubits. The first two qubits define the spin-1 and the third qubit defines the spin-1/2.
On the ibmq-lima quantum computer, we study the evolution of this system having started from the initial state projected on the positive direction of the $x$-axis. We change $Jt$ in range from 0 to $2\pi$
with step $\pi/20$. Thus for each value of $Jt$ the mean value of all components of spin-1 (\ref{meansspin1}) are measured. To obtain the information about the mean value of certain components with predefined $Jt$ the
ibmq-lima quantum computer makes 1024 shots. In Fig.~\ref{mean_val_spin1_IM} we show the time-dependence of the mean value of spin-1 with respect to $Jt$. The solid line shows the theoretical prediction
\begin{eqnarray}
\vert\langle\psi(t)\vert{\bf S}_1\vert\psi(t)\rangle\vert=\vert\cos(Jt)\vert,
\label{IMmeanspin}
\end{eqnarray}
As we can see that the experimental results agree well with theoretical ones. 

Now, let us consider the time dependence of correlation between spin-1 and spin-1/2. We measure the $xx$-component of correlations. Similarly as in the equation (\ref{correlatfunc}) we write
\begin{eqnarray}
&&\langle\psi(t)\vert S^x_1\sigma^x_2\vert\psi(t)\rangle=\langle\tilde{\psi}(t)\vert S^z_1\sigma^z_2 \vert\tilde{\psi}(t)\rangle\nonumber\\
&&=\vert\langle\tilde{\psi}(t)\vert\vert 1,1\rangle_1\vert 0\rangle_2\vert^2-\vert\langle\tilde{\psi}(t)\vert\vert 1,1\rangle_1\vert 1\rangle_2\vert^2\nonumber\\
&&-\vert\langle\tilde{\psi}(t)\vert\vert 1,-1\rangle_1\vert 0\rangle_2\vert^2+\vert\langle\tilde{\psi}(t)\vert\vert 1,-1\rangle_1\vert 1\rangle_2\vert^2,
\label{IMcorrelatfunc}
\end{eqnarray}
where $\vert\tilde{\psi}(t)\rangle=e^{\frac{\pi}{2}S^y_1}e^{\frac{\pi}{4}\sigma^y_2}\vert\psi(t)\rangle$. The analytical expression of this correlation function has the form
\begin{eqnarray}
\langle\psi(t)\vert S^x_1\sigma^x_2\vert\psi(t)\rangle=\cos(Jt).
\label{IMcorrelatfunc2}
\end{eqnarray}
In Fig.~\ref{corrIM} we present the results obtained on the ibmq-lima quantum computer. Despite the fact that we measured all three qubits, here we also obtained good agreement with the theoretical prediction.

We estimate the maximal possible value of the relative error for the results shown in Fig.~\ref{meanvalspin1corrIM}. This error can be represented as the sum of the gates error that defines the quantum circuit, the readout error,
and the error of counting statistics, which is inversely proportional to the square root of the shots (in our case $1/32=3.125\%$). To implement the Ising model shown in Fig.~\ref{schemeIM} the ibmq-lima uses 20 single-qubit gates
and 4 controlled-not gates. The average single-qubit and controlled-not gates error on this quantum device are $0.047\%$ and $1.168\%$, respectively. Thus, the total error that appears due to gates is equal
to $20\cdot 0.047\%+4\cdot 1.168\%=5.612\%$. The average readout error of one qubit is $2.63\%$. In the case of the measurement of the mean value of spin-1 (Fig.~\ref{mean_val_spin1_IM}) the maximal possible error appears
in the case of $\vert\langle S_1\rangle\vert=1$. In this case, we measure two qubits, therefore the total error consists of gates error, the sum of readout errors of two qubits, and the error of counting statistics $5.612\%+2\cdot 2.63\%+1/32=3.125\%\approx 14\%$.
In the case of the correlation between spins, it should be measured in three qubits. In turn, we take into account the error of the third qubit. Therefore the maximal possible error appears in case of $\vert\langle S_1^x\sigma_2^x\rangle\vert=1$
is equal to$5.612\%+3\cdot 2.63\%+1/32=3.125\%\approx 16.63\%$. These errors are defined by the bars in Fig.~\ref{meanvalspin1corrIM}. As we can see the results obtained on the quantum computer and theoretical prediction are in the area of error bars overlap.

Note that modeling of the systems described by more complicated Hamiltonians such as the Heisenberg model or anisotropic Heisenberg model requires the application of the Suzuki-Trotter decomposition \cite{trotter1959,suzuki2005}.
In turn, this leads to the increase of the quantum gates in the circuit and a fast accumulation of errors.

\section{Simulating spin-s on a quantum computer \label{spins}}

To simulate the spin of value $s$ a set of $N=2s$ spins $1/2$ can be used. Then the spin operator takes the form
\begin{eqnarray}
{\bf S}=\frac{1}{2}\sum_{i=1}^N{\boldsymbol \sigma_i},
\label{spinoperator}
\end{eqnarray}
where ${\boldsymbol \sigma_i}=I\otimes I\otimes\ldots {\boldsymbol \sigma}\otimes \ldots \otimes I$ is the Pauli operator which corresponds to the $i$th spin-$1/2$.
Thus, eigenvalue equations (\ref{condition1}), (\ref{condition2}) for operator (\ref{spinoperator}) take the form
\begin{eqnarray}
&&\frac{1}{4}\left(3NI+\sum_{i\neq j}^N\left(\sigma_i^x\sigma_j^x+\sigma_i^y\sigma_j^y+\sigma_i^z\sigma_j^z\right)\right)\vert s, m\rangle\nonumber\\
&& = s(s+1)\vert s, m\rangle,\label{spinoperatoreigeneq1}\\
&&\frac{1}{2}\sum_{i}^N\sigma_i^z\vert s, m\rangle = m\vert s, m\rangle,
\label{spinoperatoreigeneq2}
\end{eqnarray}
The eigenstates $\vert s, m\rangle$ can be expressed as follows
\begin{eqnarray}
&&\vert s, m\rangle= \frac{1}{\sqrt{C_N^{s-m}}}\nonumber\\
&&\sum_{i_1<i_2<\ldots<i_{(s-m)}}^N\sigma_{i_1}^x\sigma_{i_2}^x\ldots\sigma_{i_{(s-m)}}^x\vert 0\rangle_1\vert 0\rangle_2 \ldots \vert 0\rangle_N,
\label{szeigenstates}
\end{eqnarray}
where $C_N^{s-m}=\frac{N!}{(s-m)!(N-s+m)!}$. Here we use the fact that $\sigma^x\vert 0\rangle=\vert 1\rangle$. Note that in the case $m=s$ we have the state $\vert s, s\rangle=\vert 0\rangle_1\vert 0\rangle_2 \ldots \vert 0\rangle_N$.
It is easy to convince oneself that states (\ref{szeigenstates}) satisfy equations (\ref{spinoperatoreigeneq1}), (\ref{spinoperatoreigeneq2}). Really, states (\ref{szeigenstates}) are eigenstates of operator $\sigma_i^x\sigma_j^x+\sigma_i^y\sigma_j^y+\sigma_i^z\sigma_j^z$
with eigenvalue $1$. We have $N(N-1)$ such operators in equation (\ref{spinoperatoreigeneq1}). Therefore, after the action of the square of the operator (\ref{spinoperator}) on the state (\ref{szeigenstates}) we obtain this state
with factor $1/4(3N+N(N-1))=N/2(N/2+1)=s(s+1)$. In the case with equation (\ref{spinoperatoreigeneq2}), each term in state (\ref{szeigenstates}) is the eigenstate of operator $\frac{1}{2}\sum_{i}^N\sigma_i^z$
with eigenvalue $m$. Note, that in paper \cite{Sinha2003} it was mentioned realization of spin-1 and spin-3/2 with the help of two and three spin-1/2, respectively. The states which the authors proposed for simulation
of the Hilbert space are the particular cases ($s=1$, $s=3/2$) of states defined by expression (\ref{szeigenstates}).

Thus, there are $2s+1$ of states that define the Hilbert space of spin-$s$. States (\ref{szeigenstates}) of $N=2s$ qubits can be considered to simulate the behaviour of spin-$s$ on a quantum computer.
Using these states an arbitrary state of spin-$s$ can be expressed as follows
\begin{eqnarray}
\vert\psi_s\rangle=\sum_{m=-s}^s C_m\vert s,m\rangle,
\label{arbstatespins}
\end{eqnarray}
where $C_m$ are the complex parameters which satisfy the normalization condition $\sum_{m=-s}^s\vert C_m\vert^2$ and $\vert s,m\rangle$ are defined by expression (\ref{szeigenstates}).
As we mentioned earlier the quantum computer provides measurements on the basis of states of qubits. In the case of spin-$s$, there are $2^N$ basis states. The amplitudes defined state (\ref{arbstatespins}) are related with probabilities measured on the quantum computer in the following way
\begin{eqnarray}
\vert C_m\vert^2=\sum_{k=1}^{C_N^{s-m}}P_{m,k},
\label{relationsampl1}
\end{eqnarray}
where $P_{m,k}$ are the probabilities which correspond to the measurements on basis vectors which define state (\ref{szeigenstates}). Then the mean value of $S^z$ operator in state $\vert\psi\rangle$ can be expressed as follows
\begin{eqnarray}
\langle\psi\vert S^z\vert\psi\rangle=\sum_{m=-s}^{s}m\vert\langle\psi\vert s,m\rangle\vert^2=\sum_{m=-s}^{s}m\vert C_m\vert^2.
\label{meanszs}
\end{eqnarray}
The $zz$-correlation between $i$th and $j$th spins has the form
\begin{eqnarray}
\langle\psi\vert S^z_iS^z_j\vert\psi\rangle=\sum_{m=-s}^{s}\sum_{m'=-s}^{s}mm'\vert\langle\psi\vert\vert s,m\rangle_i\vert s,m'\rangle_j\vert^2.\nonumber\\
\label{corrszs}
\end{eqnarray}
Similarly as in the case of spin-1 (Sec.~\ref{spin1}), using rotating representations (\ref{representationofspinoperator}), we can measure other means and correlations of the spin-s operator.

\section{Conclusions \label{conc}}

We have considered a method that allows one to simulate the high spins by the qubits of a quantum computer. This method is based on the fact that the operator of spin-$s$ can be generated by 
some compositions of spin-$1/2$ operators, which satisfy the Lie algebra for the spin-s operator and equations (\ref{condition1}), (\ref{condition2}). In this way, we have constructed the spin operator and basis states
which define the Hilbert space of the spin-$1$. Protocols that allow one to measure the mean value of spin-s and correlations between two certain spins have been proposed. We have used these protocols
to simulate the evolution of spin-1 in the magnetic field on the ibmq-santiago quantum computer and the evolution of two interacting spins described by the Ising model on the ibmq-lima quantum computer. We have measured the mean value of spin
and correlations between spins. We have obtained a good agreement between experimental results with theoretical predictions. This means that our method can be used for modeling high-spin systems on a quantum computer.

\section{Acknowledgements}

This work was supported by Project 77/02.2020 from National Research Foundation of Ukraine.
%We are grateful to Prof. Andrij Rovenchak for helpful advices.
%This work was supported by Project FF-83F (No. 0119U002203)
%from the Ministry of Education and Science of Ukraine.

{}

\end{document}